# The gas-phase reaction of $NH_2$ with formaldehyde ($CH_2O$) is not a source of formamide ($NH_2CHO$) in interstellar environments

Kevin M. Douglas[a*], Daniel Lucas[a], Catherine Walsh[b], Niclas A. West[a], Mark A. Blitz[a,c], Dwayne E. Heard[a*]

[a]*School of Chemistry, University of Leeds, Leeds, LS2 9JT, UK*

[b]*School of Physics and Astronomy, University of Leeds, Leeds, LS2 9JT, UK*

[c]*National Centre for Atmospheric Science (NCAS), University of Leeds, Leeds, LS2 9JT, UK*

[*]corresponding authors. Email: k.m.douglas@leeds.ac.uk; d.e.heard@leeds.ac.uk

**Abstract**

The first experimental study of the low-temperature kinetics of the gas-phase reaction of $NH_2$ with formaldehyde ($CH_2O$) has been performed. This reaction has previously been suggested as a source of formamide ($NH_2CHO$) in interstellar environments. A pulsed Laval nozzle equipped with laser-flash photolysis and laser-induced fluorescence spectroscopy was used to create and monitor the temporal decay of $NH_2$ in the presence of $CH_2O$. No loss of $NH_2$ could be observed via reaction with $CH_2O$ and we place an upper-limit on the rate coefficient of $<6\times10^{-12}$ $cm^3$ $molecule^{-1}$ $s^{-1}$ at 34K. *Ab initio* calculations of the potential energy surface were combined with RRKM calculations to predict a rate coefficient of $6.2\times10^{-14}$ $cm^3$ $molecule^{-1}$ $s^{-1}$ at 35K, consistent with the experimental results. The presence of a significant barrier, 18 kJ $mol^{-1}$, for the formation of formamide as a product, means that only the H-abstraction channel producing $NH_3$ + CHO, in which the transfer of an H-atom can occur by quantum mechanical tunnelling through a 23 kJ $mol^{-1}$ barrier, is open at low temperatures. These results are in contrast with a recent theoretical study which suggested that the reaction could proceed without a barrier and was therefore a viable route to gas-phase formamide formation. The calculated rate coefficients were used in an astrochemical model which demonstrated that this reaction produces only negligible amounts of gas-phase formamide under interstellar and circumstellar conditions. The reaction of $NH_2$ with $CH_2O$ is therefore not an important source of formamide at low temperatures in interstellar environments.

## 1. Introduction

A major open question in astrochemistry concerns the mechanisms for the formation of complex organic molecules (COMs), with particular interest in those molecules that may play a role in prebiotic chemistry. Formamide ($NH_2CHO$), is one such molecule. Being the smallest molecule to contain the peptide bond (NH-C=O), the type of bond that plays a key role in linking amino acids into peptide chains and proteins, it contains all the components necessary for the formation of nucleic polymers under prebiotic conditions (Saladino et al. 2012). Formamide was first detected in the high-mass star-forming region (SFR) Sgr B2 by Rubin et al. (1971). Since then it has been observed in several other high-mass SFRs, as well as other astrochemical regions such as hot corinos and in protostellar shocks (Lopez-Sepulcre et al. 2019). The detection of formamide in comets (Biver et al. 2014; Bockelee-Morvan et al. 1997; Goesmann et al. 2015) also raises the question of whether it may have been exogenously delivered onto planetary bodies such as the early Earth. Despite the apparent ubiquitous nature of formamide in the interstellar medium, the mechanisms for forming it are still not fully understood, and whether it is mostly formed via gas-phase or grain surface chemistry is hotly debated (Codella et al. 2017).

Both theoretical and experimental studies have investigated the formation of formamide on the surface of interstellar dust grains or in icy mantles. Many of these experiments indicate that formamide is relatively easily produced in the solid phase by the processing of ices containing H, N, C, and O precursors by a range of energy inputs (UV or energetic electron or ion impact; see Lopez-Sepulcre et al. (2019). Despite this, the refractory nature of formamide means that higher temperatures are required for its desorption into the gas phase when compared to other COMs commonly detected in the ISM (Dulieu et al. 2019). Furthermore, many of the energetic processes required for forming formamide are also destructive, with Brucato et al. (2006) indicating only around 20 % of frozen formamide molecules irradiated in the dense ISM are able to survive on a timescale of $10^8$ years. Several studies have looked at the formation of formamide *via* ion-molecule reactions in the gas-phase; however, of all the reactions considered, none of them were found to be a possible route to interstellar formamide, either due to high energy barriers, or because they favour other product channels (Redondo et al. 2014a; Redondo et al. 2014b; Spezia et al. 2016).

Gas-phase formation of formamide *via* the neutral-neutral reaction of the amidogen radical, $NH_2$, and formaldehyde ($CH_2O$, ubiquitously found in space), has also been suggested as a viable route theoretically (Reaction R1b; (Barone et al. 2015; Skouteris et al. 2017; Vazart et al. 2016). For the reaction between $NH_2 + CH_2O$, there are two primary exothermic product channels; a hydrogen-abstraction channel in which the $NH_2$ abstracts an H atom from formaldehyde to produce ammonia, $NH_3$, and the formyl radical, CHO, (reaction R1a), and an addition-elimination channel in which the $NH_2$ first attacks the C of the formaldehyde to form a bound adduct, which then goes on to eliminate an H atom and produce formamide (reaction R1b). A third exothermic product channel to E-methanimidic acid + H first requires the formation of the bound adduct as for reaction R1b, followed by an extensive rearrangement over barriers of 50 kJ mol$^{-1}$ or more (Vazart et al. 2016), and hence is not considered in this study. There have been several previous theoretical studies investigating the $NH_2$ + formaldehyde potential energy surface (PES); however these studies have only considered either the H-abstraction channel or the formamide + H reaction channel, with no studies considering the full surface. Li and Lü (2002) calculated the minimum energy pathway (MEP) of the H-abstraction channel at the MP2/6-311+G(d,p) level of theory, as well as performing single-point energy refinements of the stationary points at the G2//MP2/6-311+G(d,p) level of theory. Using this surface, Li and Lü calculated rate coefficients for the H-abstraction channel using both conventional transition state theory and canonical variational transition state theory, over the temperature range of 250 – 1500 K. Depending on the level of theory and method used, Li and Lü predict the rate of H-abstraction at 250 K to be very slow, ranging between 5

× 10$^{-20}$ and 1 × 10$^{-17}$ cm$^3$ molecule$^{-1}$ s$^{-1}$. Barone et al. (2015) mapped out the PES of the H + formamide channel at the B2LYP/m-aug-cc-pVTZ level of theory, as well as computing more accurate electronic energies using the complete basis set QB3 method (Montgomery et al. 2000; Ochterski et al. 1996). Using an in-house code they have also calculated rate coefficients for the H + formamide product channel over the temperature range of 10 – 300 K, predicting the reaction to have in inverse temperature dependence, with the rate of formamide production rising from ~ 2 × 10$^{-12}$ cm$^3$ molecule$^{-1}$ s$^{-1}$ at 300 K, up to ~ 3 × 10$^{-10}$ cm$^3$ molecule$^{-1}$ s$^{-1}$ at 10 K. It should be noted that these fast rates were calculated using a PES that has no barrier to adduct formation. The PES of the H + formamide channel was revisited in two further papers (Skouteris et al. 2017; Vazart et al. 2016), both of which also omit the barrier to adduct formation and predict a fast rate coefficient for the formation of the formamide product. Finally, Song and Kästner (2016) optimized the reactants and the TS for forming the bound adduct at the M06-2X/def2-TZVP level of theory, and calculated single-point energies and vibrational frequencies at the UCCSD(T)-F12/cc-pVTZ-F12 level. In agreement with Vazart et al. (2016), they find an almost submerged barrier to adduct formation, +2.7 kJ mol$^{-1}$ compared to the separated reactants. However, including the ZPE increases the barrier height to +17.8 kJ mol-1; as such, they conclude that reaction R1b does not play a significant role in formation of formamide, calculating a rate coefficient of ~ 5 × 10$^{-22}$ cm$^3$ molecule$^{-1}$ s$^{-1}$ for reaction R1b at 100 K. This rate is over ten orders of magnitude smaller than the rate predicted by Barone et al. (2015) and Skouteris et al. (2017) at 100 K, which are at present the rates listed by the Kinetic Database for Astrochemistry (KIDA) (Wakelam et al. 2012) for reaction R1b.

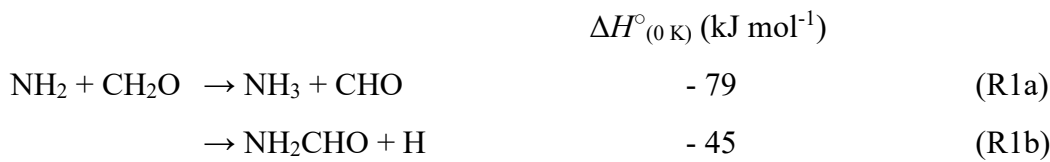

$\Delta H°_{(0\ K)}$ (kJ mol$^{-1}$)

| | | |
|---|---|---|
| NH$_2$ + CH$_2$O → NH$_3$ + CHO | - 79 | (R1a) |
| → NH$_2$CHO + H | - 45 | (R1b) |

In this paper, we present the first experimental results into the reaction between NH$_2$ and CH$_2$O, using a pulsed laser photolysis-laser induced fluorescence (PLP-LIF) technique coupled with a Laval nozzle to achieve the low temperatures relevant to the ISM. We also present results from a theoretical investigation into the reaction, in which we predict rate coefficients for the two product channels over the temperature range 10 – 350 K. These rate coefficients are then incorporated into an astrochemical model representative of the L1157-B2 shocked region and the circumstellar environment around IRAS 16293, following the same model set-up and methodology as presented in Barone et al. (2015).

## 2. Methodology

### 2.1 Experimental Study

The low temperature kinetics of the reaction of NH$_2$ with CH$_2$O were measured using a PLP-LIF technique coupled with a Laval nozzle expansion, a method that has been described in detail previously (Caravan et al. 2015; Gomez Martin et al. 2014; Shannon et al. 2013; Taylor et al. 2008). As such, only a brief overview is given here.

The reagent (NH$_3$ (99.98 %), BOC), CH$_4$ (99.995 %, BOC) and bath gases (He (99.9995 %), N$_2$ (99.9995 %), Ar (99.9995 %); BOC) were combined in a mixing manifold using calibrated mass flow controllers (MFCs; MKS Instruments), prior to entering a 2 L gas ballast tank. The CH$_2$O reagent was introduced as a dilute mixture in bath gas. The CH$_2$O mixtures were prepared in cylinders by heating paraformaldehyde (Sigma-Aldrich, 95 %), using the method as described in West et al. (2019). The formaldehyde concentration used in each experiment was measured directly by UV absorption spectroscopy, details of which are given in the SI. Following the gas ballast, the reaction mixture was introduced to a 1 cm$^3$ stainless steel reservoir *via* two pulsed solenoid valves (Parker 9 series), fired at a repetition rate of either 5 or 10 Hz, with a pulse duration of around 10 ms. Each pulse of gas underwent a controlled expansion through a convergent-divergent shaped Laval nozzle into a low-pressure stainless-steel cylindrical chamber (~ 775 mm length by 240 mm diameter), resulting in a thermalized low temperature gas flow. A range of nozzles were employed during the experiments to achieve flow temperatures of between 34 and 72 K. The temperature and density profile of the flows were characterized by impact pressure measurements, and the temperature of several of the jets confirmed by rotationally resolved LIF spectroscopy (Douglas et al. 2018; West et al. 2019).

NH$_2$ radicals were generated from the PLP of NH$_3$ (R3) at 213 nm (Reaction R3) by the 5$^{th}$ harmonic of a Nd:YAG laser (Quantel Q-Smart 850), with a typical pulse energy of ~10 mJ. NH$_2$ radicals were observed by time-resolved LIF spectroscopy, probing the $A\,^2A_1$ (0,10,0) ← $X\,^2B_1$ (0,0,0) transition near 597.6 nm (Copeland et al. 1985; Donnelly et al. 1979) using the output of Nd:YAG pumped dye laser (a Quantel Q-smart 850 pumping a Sirah Cobra-Stretch). The non-resonant fluorescence at ~ 620 nm was collected *via* a series of lenses through an optical filter (Semrock Brightline interference filter, $\lambda_{max}$ = 620 nm, fwhm = 14 nm), and observed by a temporally gated channel photomultiplier (CPM; PerkinElmer C1952P), mounted at 90° to both laser beams. The signal from the CPM was recorded using a digital oscilloscope (LeCroy Waverunner LT264), and sent to a computer using a custom LabVIEW program. The temporal evolution of the LIF signal was recorded by varying the time delay between the photolysis and probe lasers. A typical time-resolved LIF profile (inset Figure 1) consisted of 110 delay steps and resulted from the average or between 10 and 20 individual delay scans.

NH$_3$ + $h\nu$ → NH$_2$ + H  (R3)

For experiments monitoring formaldehyde dimerization, formaldehyde was observed by probing the $A\,^1A_2$ ($4^1_0$) ← $X\,^1A_1$ ($0^0_0$) transition near 353 nm,(Burkert et al. 2000; Clouthier and Ramsay 1983) using the frequency doubled output of a Nd:YAG pumped dye laser (same system as described above with a BBO doubling crystal). The non-resonant fluorescence at $\lambda$ > 390 nm was discriminated using a long-pass Perspex filter.

2.2   *Theoretical Calculations*

All electronic structure calculations were carried out using the Gaussian 09 suite of programs (Frisch et al. 2016). The stationary points on the full NH$_2$ + formaldehyde surface were mapped out at the B3LYP/6-311+G(2d,p) level. The structures of the stationary points were further optimized at the M062X/aug-cc-pvtz level, from which rotational constants, harmonic vibrational frequencies, and ZPEs were obtained. High-performance single point energies were also calculated at the CCSD(T) level using the M062X structures. The single point energies were extrapolated to the complete basis set limit (CBS) using the aug-cc-pVXZ basis sets (X = 3, 4, 5) and a mixed Gaussian/exponential extrapolation scheme as proposed by Peterson et al. (1994). RRKM calculations were performed using the Master Equation Solver for Multi-Energy well Reactions (MESMER) program (Glowacki et al. 2012).

## 3. Results

Typical $NH_2$ LIF temporal profiles produced following the PLP of $NH_3$ in the presence of $CH_2O$ can be seen in the inset of Figure 1. In these experiments, the absolute $NH_2$ LIF signal decreased with increasing $CH_2O$. We attribute this to the reaction of the basic $NH_3$ precursor with the slightly acidic $CH_2O$, as discussed above. We are unable to tell whether this reaction was occurring in the ballast chamber, in the pre-expansion reservoir, or in the low temperature flow (or any combination thereof). However, we were still able to observe good $NH_2$ signal, indicating that even at the highest $CH_2O$ used, sufficient $NH_3$ remained to be photolyzed. No accounting of any loss of $CH_2O$ *via* this reaction was required, as these experiments used $CH_2O$ concentrations significantly higher than that of $NH_3$. As can be seen from Figure 1, there is an initial growth of the $NH_2$ signal, with very little instant signal observed. This growth in $NH_2$ ($v = 0$) signal is due to collisional relaxation of vibrationally excited $NH_2$, as has been observed in previous studies (Yamasaki et al. 2002a; Yamasaki et al. 2002b). As kinetics measurements are limited by the dynamic time of the low temperature jets, it was important that this relaxation was as efficient as possible, in order to maximise the time in which we could observe the loss of $NH_2$ ($v = 0$). To this end, $CH_4$, which we have shown to be an efficient in relaxing $NH_2$ ($v > 0$), was added to our gas flows.

The $NH_2$ traces were fitted satisfactorily with a biexponential growth and loss (solid lines inset Figure 1), yielding pseudo-first-order loss rates, $k'_{obs}$, from which bimolecular plots of $k'_{obs}$ *vs* [$CH_2O$] were produced (Figure 1). Typically, $NH_2$ traces were collected at 8 or more different $CH_2O$ concentrations. It is important to determine the maximum concentration of $CH_2O$ we were able to add to our low temperature flows before significant amounts of $CH_2O$ dimers began to form, as this could result in an error in the reported rate coefficient. Details of experiments conducted monitoring $CH_2O$ dimerization are given in the SI. We found that significant complex formation occurred at much lower [$CH_2O$] when using either Ar or $N_2$ as a bath gas as compared to He, presumably due to both Ar and $N_2$ being better third bodies than He. In the He flow at ~ 34 K, formaldehyde dimerization occurred around a concentration of ~ $3 \times 10^{14}$ molecule cm$^{-3}$, whereas in the $N_2$ flow at ~ 72 K, and the Ar flow at ~ 40 K, dimerization occurred at around $5 \times 10^{13}$ and $3 \times 10^{13}$ molecule cm$^{-3}$ respectively. Dimerization experiments in He were carried out both with and without $CH_4$ and $NH_3$ present, and no change in the concentration at which significant $CH_2O$ dimerization occurred was observed.

As can be seen from Figure 1, no increase in the removal of $NH_2$ was observed as the formaldehyde concentration was increased, indicating that this reaction is slower than we are able to measure in our experiments. We are, however, able to put an upper limit on the rate of removal of NH$_2$ by CH$_2$O. For each experiment performed, we have determined the minimum change in $k'_{obs}$ that would be clearly observable in our experiment. This was taken as 2 × the standard deviation of the $k'_{obs}$ values over the whole [CH$_2$O] range obtained for a particular experiment, i.e. 2 × the variance of the $k'_{obs}$ values. By dividing this by the maximum concentration of formaldehyde added (up to a maximum of 3 × 10$^{14}$ molecule cm$^{-3}$, the concentration at which dimers begin to form in He at ~ 34 K), an upper limit on the rate of removal of NH$_2$ with CH$_2$O was calculated. Table 1 provides the upper limits calculated for several experiments carried out. Taking the average of the three lowest values give an upper limit of 6 × 10$^{-12}$ cm$^3$ molecule$^{-1}$ s$^{-1}$ at $T$ = 34 K. This upper limit was determined using He as a bath gas and with our lowest temperature nozzle. We would be unable to improve on this upper limit using higher temperature nozzles, or different bath gases, for two reasons. The first is due to the dynamic times available in our experiments. Our lowest temperature nozzle also provides the longest possible dynamic time in our experiments. Our higher temperature nozzles have shorter dynamic times. With a shorter time in which to observe the loss of NH$_2$, we would expect greater scatter in the observed removal rates of NH$_2$ (i.e. the $k'_{obs}$ values), and as such a larger upper limit on the NH$_2$ + CH$_2$O removal rate. When using Ar and N$_2$ as a bath gas, we are able to get significantly longer dynamic times than when using He. However, here any improvement in the scatter of $k'_{obs}$ from the increased dynamic time would be outweighed by the significant reduction in the maximum amount of formaldehyde we are able to introduce in an Ar or N$_2$ flow without CH$_2$O dimers forming. This is approximately 10 times less than we are able to add when using He as a bath gas. Therefore, the upper limit we quote of 6 × 10$^{-12}$ cm$^3$ molecule$^{-1}$ s$^{-1}$ at $T$ = 34 K is the best limit we are able to provide in our experiments. To the best of our knowledge, there are no other studies which have reported a rate coefficient for the removal of NH$_2$ with CH$_2$O, with our upper limit being the first. It should be noted that the experimental upper limit relates to the total removal of NH$_2$ by CH$_2$O, i.e. the sum of both product channels. However, as our theoretical results show that the formamide + H product channel is effectively turned off below ~ 100 K (see below), the upper limit relates solely to the H-abstraction channel. However, this upper limit should not be taken at the rate coefficient at $T$ = 34 K, as discussed below we calculate the rate at $T$ = 35 K to be significantly smaller (see Figure 3).

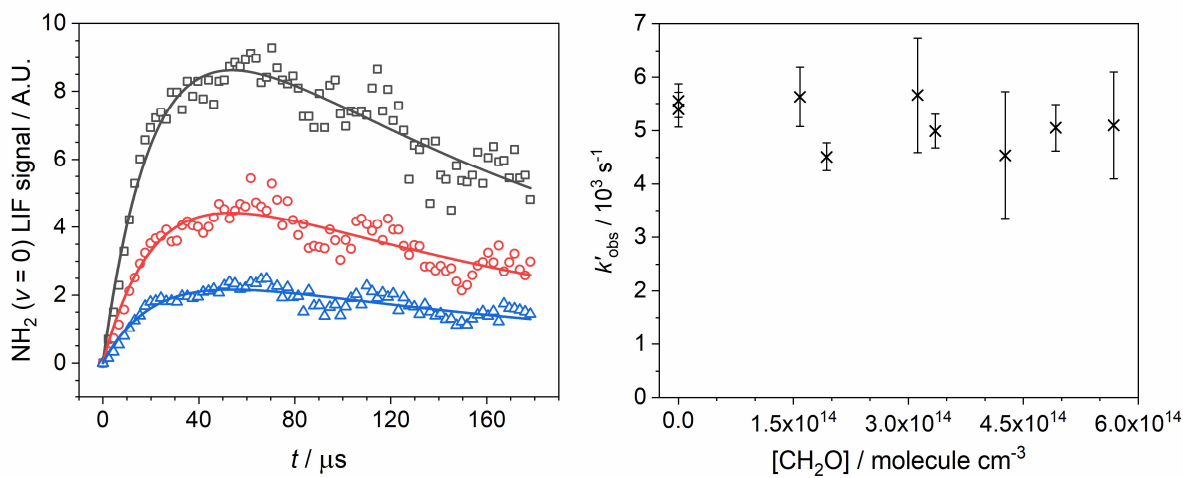

**Figure 1.** Left: NH$_2$ ($v$ = 0) traces (an average of between 10 and 20 individual delay scans) collected at $T$ = 34.1 K, a total He density = 4.12 × 10$^{16}$ molecule cm$^{-3}$, and [CH$_2$O] of 0, 1.6, and 3.1 × 10$^{14}$ molecule cm$^{-3}$ (black squares, red circles, and blue triangles respectively). Solid lines are the least-squares fitting of a biexponential to the traces from which $k'_{obs}$ is obtained. Right: Bimolecular plot of $k'_{obs}$ vs [CH$_2$O] at $T$ = 34.1 K, showing no change in the removal of NH$_2$ ($v$ = 0) with formaldehyde (errors at the 1 σ level).

**Table 1.** Calculated upper limits on the rate coefficients for the reaction of NH$_2$ + CH$_2$O and relevant experimental conditions.

| Bath Gas | $T^a$ / K | $N_{total}^a$ / 10$^{16}$ cm$^{-3}$ | 2 × σ($k'_{obs}$)$^b$ / s$^{-1}$ | [CH$_2$O]$_{max}$ / 10$^{14}$ molecule cm$^{-3}$ | $k_{1\ max}$ / 10$^{-12}$ cm$^3$ molecule$^{-1}$ s$^{-1}$ |
|---|---|---|---|---|---|
| He | 33.9 ± 2.0 | 6.21 ± 0.51 | 1700 | 3.4 | 5.7$^c$ |
|  | 34.1 ± 2.8 | 4.12 ± 0.50 | 1300 | 2.0 | 6.5 |
|  | 34.1 ± 2.8 | 4.12 ± 0.50 | 1750 | 5.7 | 5.8$^c$ |
|  | 34.1 ± 2.8 | 4.12 ± 0.50 | 750 | 0.42 | 18 |
|  | 34.1 ± 2.8 | 4.12 ± 0.50 | 2400 | 3.3 | 8.0$^c$ |

$^a$Uncertainties in each value of $T$ and $N_{total}$ are ± 1σ (the standard deviation) of the measured temperature and density along the axis of the Laval expansion. $^b$The 2 × σ($k'_{obs}$) values were determined by taking the standard deviation of the $k'_{obs}$ values over the whole [CH$_2$O] range for a particular experiment, and multiplying by 2. $^c k_{1\ max}$ values calculated using [CH$_2$O]$_{max}$ value of 3.0 × 10$^{14}$ molecule cm$^{-3}$, as this is the maximum CH$_2$O that can be added before significant CH$_2$O dimers begin to form.

## 4. Discussion

### 4.1 Theoretical Calculations

A schematic of the full PES for the reaction between NH$_2$ and CH$_2$O can be seen in Figure 2. The energies given are CCSD(T) extrapolated energies and include ZPEs calculated at the M062x level. The full molecular properties of the stationary points can be found in Table S1. As can be seen from Figure 2, the reaction may initially proceed *via* the formation of one of two pre-reaction complexes (PRCs). The hydrogen-bonded (HB) PRC is linked to the formation of the products NH$_3$ + CHO *via* TS1; as this involves the transfer of an H-atom, quantum mechanical tunnelling is likely to play a role at low temperatures. The van-der-Waals (vW) PRC is first linked to the formation of bound adduct *via* TS2, from which the elimination of an H-atom *via* TS3 will produce formamide + H.

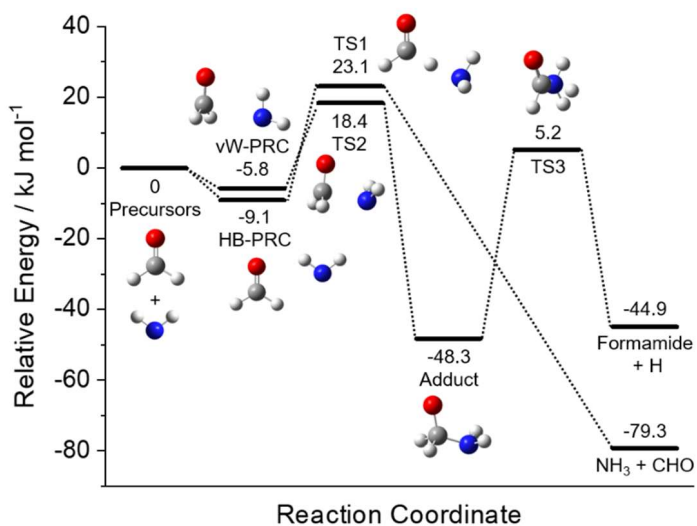

**Fig 2.** Potential energy surface for $NH_2$ + $CH_2O$ determined at the CCSD(T)//M062X-aug-cc-pVTZ level of theory.

Table S2 compares our calculated energies for the stationary points on the PES with those available in the literature. Our calculated ZPE corrected energies for the barrier height of TS1 (for H-abstraction) and the heat of reaction for forming $NH_3$ + CHO are in good agreement with those calculated by Li and Lü (2002), being within 1.7 kJ mol$^{-1}$ of each other. Good agreement is also observed between the barrier height of TS2 (from the vW-PRC to the adduct) calculated in this study to those calculated by Vazart et al. (2016) and by Song and Kästner (2016), with ours being 0.6 kJ mol$^{-1}$ lower and higher respectively. Our ZPE corrected energies for the adduct, TS3 (from the adduct to formamide + H), and for the products formamide + H are also broadly in agreement with those calculated previously, being within 5 kJ mol$^{-1}$ to that from Barone et al. (2015), and within 2.3 kJ mol$^{-1}$ to a more recent study (Vazart et al. 2016). It should be noted that in the studies by Li and Lü (2002) and Song and Kästner (2016), no PRCs are reported, and that while both Barone et al. (2015) and Skouteris et al. (2017) do mention the presence of PRCs on the PES, no details of them are given. Vazart et al. (2016) does report on the presence of both the HB-PRC and the vW-PRC, for which our calculated electronic energies are in excelled agreement, lying within 0.2 kJ mol$^{-1}$ of each other; however, as they do not consider the H-abstraction channel, they suggest both PRCs lead to TS2. We have performed intrinsic reaction coordinate (IRC) calculations of all three transition states, which indicate that the HB-PRC is linked to the H-abstraction TS (TS1), while the vW-PRC is linked to the TS leading to adduct formation (TS2). This discrepancy does not affect the calculated rate coefficients reported by Vazart et al. (2016), who, like Barone et al. (2015) and Skouteris et al. (2017), omit both the PRCs and the barrier to adduct formation (TS2) from their surface. Barone et al. (2015) give no reason for excluding TS2 from their PES, however they do note that it lies below the reactants energy at the CBS-QB3 level, but fail to mention whether this includes ZPE. Vazart et al. (2016) discuss how the electronic energy they calculate for TS2 drops close to zero when including higher excitation orders in the CCSD(T) calculation, but do not give any reason why this barrier with its substantial ZPE is excluded from their PES.

Skouteris et al. (2017) does give a reason for excluding TS2, suggesting that (i) the electronic energy of TS2 drops when including higher excitation orders in CCSD(T) calculation, and that the electronic energy will drop below the reactant level when extrapolating to the full configuration limit, and (ii) the use of the ZPE correction for the PRCs and TS2 is not warranted, as 3 of the new vibrational modes present in the structures consist of a loose stretching mode and two loose bending modes that almost constitute free rotations, and as such will be grossly overestimated. What should be noted is that the contribution to the ZPE of these 3 low frequency modes is small, and that even if these 3 frequencies were overestimated, the ZPE would only be marginally decreased. With the ZPE of TS2 (~ 15 kJ mol-1) raising the barrier to adduct formation to ~18 kJ mol$^{-1}$ above the reactants, even a substantial reduction in the ZPE would still result in a significant barrier. As such, the presence of the barrier TS2 should not be ignored.

The full $NH_2$ + $CH_2O$ PE surface was employed in the MESMER calculations, with further details of the parameters used in the calculations in the SI. The temperature dependent rate coefficients for the two product channels of reaction R1 predicted by MESMER can be seen in Figure 3, while the total rate coefficient and BR for formamide production can be seen in Table S3. Rate coefficients were calculated over the temperature range 10 – 350 K, in intervals of 5 K. The rate coefficients reported are the low-pressure limiting rate coefficients applicable to the ISM; high pressure rate coefficients that may be applicable to other environments are given in Figure S4, while figure S5 gives the calculated rate coefficients at 1 × $10^{17}$ molecule cm$^{-3}$, a pressure more comparable to those in our experimental setup. The total removal rate predicted at ~35 K is 7 × $10^{-14}$ cm$^3$ molecule$^{-1}$ s$^{-1}$, being consistent with the upper limit of 6 × $10^{-12}$ cm$^3$ molecule$^{-1}$ s$^{-1}$ determined in this study. Looking at the branching ratio between H-abstraction (R1a) and formamide production (R1b), it can be seen that channel R1b is a minor channel at all temperatures, only accounting for ~ 8 % of the total rate coefficient at 350 K, and dropping to effectively 0 % at temperatures of 130 K and below. This is due to the relatively high barrier to forming the adduct, which effectively turns off this channel at low temperatures. Song and Kästner (2016) predicted a rate of formamide production at 100 K of ~ 5 × $10^{-22}$ cm$^3$ molecule$^{-1}$ s$^{-1}$; our results are in broad agreement, with MESMER predicting a BR for R1b of 3.6 × $10^{-7}$ and a total rate coefficient of 2.4 × $10^{-15}$ cm$^3$ molecule$^{-1}$ s$^{-1}$, giving a rate coefficient for formamide production of 8.6 × $10^{-22}$ cm$^3$ molecule$^{-1}$ s$^{-1}$ at 100 K. This is in stark contrast to the results of Barone et al. (2015), who predict the rate of formamide production at 100 K to be ~ 2 × $10^{-11}$ cm$^3$ molecule$^{-1}$ s$^{-1}$, over 10 orders of magnitude faster than that predicted in this study and by Song and Kästner (2016). Additionally, Barone et al. (2015) predict the rate of formamide production at ~ 35 K to be ~ 1 × $10^{-10}$ cm$^3$ molecule$^{-1}$ s$^{-1}$, around 17 times faster than the upper limit for the total rate coefficient experimentally determined in this study, indicating that TS2 is indeed above the reactant level and should not be omitted. Looking at the rate coefficient for channel R1a, the H-abstraction channel, we can see a turnaround in the rate at around 190 K; above this temperature, we see a small positive temperature dependence, whereas below this temperature we see a strong negative temperature dependence. This sharp increase in rate at low temperature is the result of the small PRC wells before the barrier to H-abstraction, which are sufficiently long-lived at low temperatures to allow the H atom to quantum mechanically tunnel through the H-abstraction barrier (TS1) to products. This mechanism has been reported previously for a range of low temperature H-abstraction reactions involving OH and oxygenated VOCs (Blazquez et al. 2020; Caravan et al. 2015; Gomez Martin et al. 2014; Shannon et al. 2013). The H-abstraction rate coefficients calculated by Li and Lü (2002) at 250 K and above are significantly slower than those calculated in this study, and show a much stronger positive temperature dependence. This stronger temperature dependence means that while the rate coefficients reported in this study at 350 K are around 20 times faster than those of Li and Lü (2002), by 250 K our rate coefficients are around 100 times faster. The reason for this discrepancy is unclear; however the barrier height of TS1 calculated by Li and Lü is 1.5 kJ mol$^{-1}$ larger than that calculated in this study, which will result in smaller H-abstraction rate coefficients at low temperatures in the Li and Lü study. Due to the unique shape of the rate coefficient for reaction R1a *vs* temperature, we were unable to parameterise the data over the whole temperature range (10 – 350 K) using the modified Arrhenius equation. Instead, the data above and below 140 K was parameterized, giving (see green and blue solid lines Figure 3; units: cm$^3$ molecule$^{-1}$ s$^{-1}$; errors are the 1σ level of a least-squares fit to the data):

$$k_{(NH_2 + CH_2O \rightarrow NH_3 + CHO)}(10 \leq T / K \leq 140) = (9.67 \pm 0.59) \times 10^{-17} \times (T / 300)^{(-2.88 \pm 0.06)} \times \exp^{[(11.7 \pm 1.9) / T]}$$

$$k_{(NH_2 + CH_2O \rightarrow NH_3 + CHO)}(140 \leq T / K \leq 350) = (4.21 \pm 0.46) \times 10^{-18} \times (T / 300)^{(9.80 \pm 0.15)} \times \exp^{[(1839 \pm 33) / T]}$$

The temperature dependence of the rate coefficient for the formamide + H product channel was parameterized over the temperature range 110 – 350 K; below this temperature, the BR for this channel was effectively zero (see red sold line Figure 3; units: cm$^3$ molecule$^{-1}$ s$^{-1}$; errors are the 1σ level of a least-squares fit to the data):

$$k_{(NH_2 + CH_2O \rightarrow H_2NCHO + H)}(110 \leq T / K \leq 350) = (8.35 \pm 0.08) \times 10^{-14} \times \exp^{[(-1928 \pm 3) / T]}$$

These new rate coefficients will be submitted to the astochemical databases KIDA (Wakelam et al. 2012) and the UMIST Database for Astrochemistry (UdFA; McElroy et al. 2013).

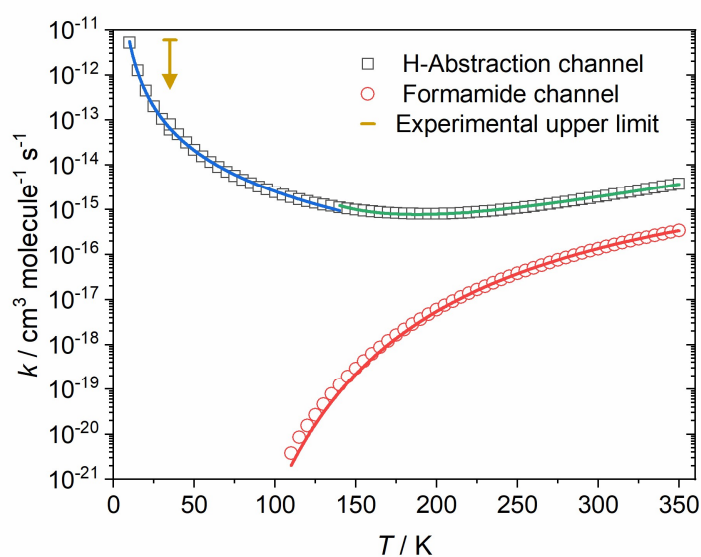

**Fig 3.** Predicted rate coefficients for the H-abstraction (black squares) and formamide + H (red circles) product channels of the $NH_2$ + $CH_2O$ reaction. The experimentally determined upper limit is shown as a yellow line, with a downward arrow below to indicate that the predicted rate is significantly smaller than this. Note that the BR for formamide production is effectively zero below ~ 100 K. Solid lines are parameterized fits to the data (see text); blue line: H-abstraction channel $T$ = 10 – 140 K; green line H-abstraction channel $T$ = 140 – 350 K; red line formamide + H channel $T$ = 110 – 350 K.

*4.2    Astrochemical Implications*

The new rate coefficients derived in the previous section were used in astrochemical models of the L1157-B2 shocked region and the cold circumstellar envelope of the IRAS 16293 protostar. Gas-phase formamide has been detected in both of these environments with a fractional abundance of $(1.1 \pm 0.2) \times 10^{-8}$ and $(3 \pm 2) \times 10^{-12}$, with respect to $H_2$, respectively (Jaber et al. 2014; Mendoza et al. 2014). The model set-ups presented in Barone et al. (2015) were also adopted here. For L1157-B2, a temperature of 70 K, a density of $10^5$ $cm^{-3}$, and a cosmic-ray ionisation rate of 3 x $10^{-16}$ $s^{-1}$, were used. The L1157-B2 model calculates the chemistry following the propagation of a shock and it is assumed that the following molecules and initial fractional abundances (with respect to total H, $n_H$) are available in the gas phase at the start of the calculation after desorption/sputtering from the icy grains: $H_2O$ ($1 \times 10^{-4}$), CO ($8 \times 10^{-5}$), $NH_3$ ($1 \times 10^{-6}$), and $CH_2O$ ($3 \times 10^{-6}$). For IRAS 16293, a temperature of 20 K, a density of $2 \times 10^6$ $cm^{-3}$, and a canonical ionisation rate of $1 \times 10^{-17}$ $s^{-1}$, were used. The initial gas composition is assumed to be in atomic form and the "EA2" set of elemental abundances from Wakelam and Herbst (2008) were adopted with additional depletion factors of 10 for C, O, and N, and 100 for heavier elements. The astrochemical model used here was from Walsh et al. (2015) and references therein, which uses the RATE12 release of the UdFA (McElroy et al. 2013) and that also includes gas-grain chemistry with grain-surface reactions and rates taken from Garrod et al. (2008). Two sets of models were run for both sources; i) a gas-phase only model, and ii) a gas-grain model.

Figure 4 (top panel) shows the fractional abundance of gas-phase formamide (with respect to $n_H$) as a function of time for the gas-phase only model of L1157-B2. The red line shows the results using the rate coefficient from Barone et al. (2015) and the blue line shows the results using the rate coefficient from this work. We show the results using a large dynamic range on the y axis to highlight the difference between the two results. The red line overlaps with the observed range (grey shaded region) from a few hundred to a few thousand years post passage of the shock; however, the new rate coefficient produces only negligible abundances of gas-phase formamide ($<< 10^{-20}$). The abundances of gas-phase formamide predicted for the gas-phase-only model for IRAS 16293 using both rate coefficients differ even more, such that the results are not able to be plotted using a meaningful scale on the y axis (i.e., the abundance is effectively 0). These results show that a gas-phase only model using the new rate coefficient for $NH_2$ + $CH_2O$ cannot reproduce the observed gas-phase abundance of formamide in these sources.

Another proposed route to gas-phase formamide is via reactions that occur on the surfaces of dust grains followed by thermal desorption in warm (> 100 K) regions, and non-thermal desorption driven by UV photons or cosmic-rays, or reactive desorption in cold (< 100 K) regions (see, e.g., Quénard et al. (2018)). These pathways include radical-radical recombination reactions such as $NH_2$ + HCO (Rimola et al. 2018), although it should be noted that many reactions included in grain-surface networks are yet to be studied neither in the laboratory nor using computational methods (see, e.g., Cuppen et al. (2017)). In Figure 4 (bottom panel) the results from a gas-grain model of the cold envelope of IRAS 16293 are presented. The fractional abundances of gas-phase $CH_2O$, $NH_2$, and $NH_2CHO$ are represented by the blue, green, and red lines respectively. Also shown in the hatched regions are the corresponding observed ranges. In contrast to the gas-phase only models, and in spite of the remaining uncertainties in grain-surface networks, the gas-grain model reproduces the observed abundances between ~ $10^5$ and ~ $10^6$ years within a factor of 3.

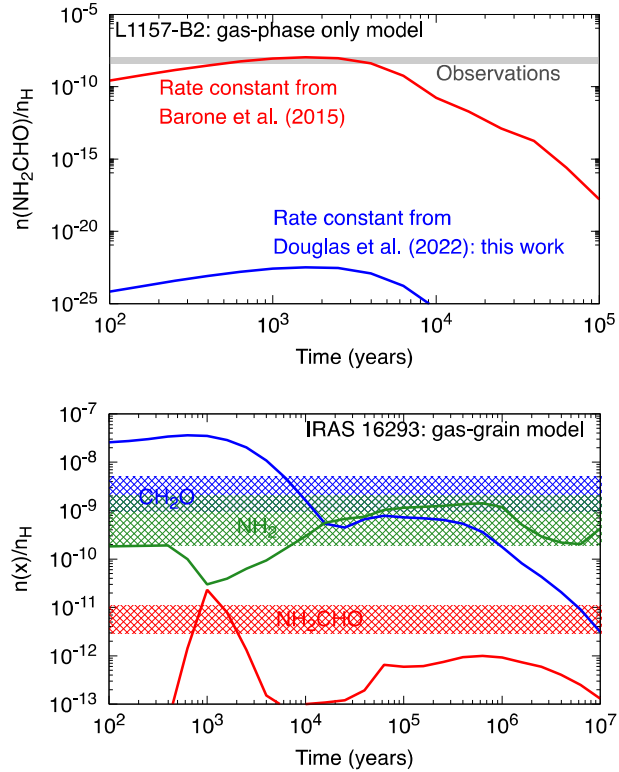

**Figure 4.** Fractional abundances with respect to total H ($n_H$) of key species as a function of time from the astrochemical models. *Top panel*: the results for gas-phase formamide using the rate coefficient for $NH_2$ + $CH_2O$ from Barone et al. (2015) in red, and this work in blue, for the gas-phase only model of L1157-B2. The observed range is shown by the grey shaded region. *Bottom panel*: the results for gas-phase $CH_2O$ (blue), $NH_2$ (green), and $NH_2CHO$ (red) for the gas-grain model of IRAS 16293. The observed ranges are shown by the coloured hatched regions.

## 5. Conclusions

The reaction between $NH_2$ and $CH_2O$ was studied using a PLP-LIF technique coupled with a Laval nozzle to achieve low temperatures. At $T = 35$ K, we were unable to observe any reaction, and only report an upper limit of $< 6 \times 10^{-12}$ cm$^3$ molecule$^{-1}$ s$^{-1}$. The full PES for the reaction was determined using electronic structure theory, and these calculations combined with RRKM theory to obtain pressure and temperature dependent rate coefficients and branching ratios. These calculations indicate that the formamide product channel is a minor channel at all temperatures, and is effectively zero below ~ 100 K. Although the H-abstraction channel dominates at all temperatures, it is also relatively slow ($2.44 \times 10^{-15}$ cm$^3$ molecule$^{-1}$ s$^{-1}$ at 100 K), only speeding up at very low temperatures when quantum mechanical tunnelling becomes efficient. These results are in stark contrast to several previous studies, in which a fast rate for formamide production is achieved by omitting the significant barrier to adduct formation. The new rate coefficients were put into astrochemical models of the L1157-B2 shock and the cold circumstellar envelope of IRAS 16293. The results show that inclusion of the new rate coefficient for the $NH_2 + CH_2O$ reaction produces negligible abundances of gas-phase formamide. On the other hand, a gas-grain model of IRAS 16293, that includes grain-surface formation pathways for formamide, can reproduce the observed abundances of $CH_2O$, $NH_2$ and $NH_2CHO$ within a factor of 3, between $10^5$ and $10^6$ years.

**Acknowledgements**

This study was supported by funding from the UK Science and Technology Facilities Council (grant ST/T000287/1). CW acknowledges support from the University of Leeds, STFC, and UKRI (grant numbers ST/T000287/1, MR/T040726/1)

\ appendix

**Supporting Information**

The gas-phase reaction of NH$_2$ with formaldehyde (CH$_2$O) is not a source of formamide (NH$_2$CHO) in interstellar environments

**UV absorption spectroscopy setup**

The formaldehyde concentration used in each experiment was measured directly by UV absorption spectroscopy, with a 1 m length absorption cell located after the mixing manifold and prior to the gas ballast. The light source was a UVB lamp (Exo Terra UVB200) with continuous output between ~ 290 and 350 nm. Absorption spectra were collected using a UV-Vis spectrometer (Ocean Optics, HR4000CG-UV-NIR) with 0.75 nm resolution and a 2 s integration window. Four spectra traces were averaged to collect an averaged spectrum from which the concentration of CH$_2$O could be determined. The pressure of the gas mixture in the absorption cell, as measured by a capacitance manometer (Baratron MKS Instruments, 0 – 5000 Torr), and was typically around 1000 Torr, equal to the pressure in the gas ballast. To fit the averaged spectra, a least-squares minimization analysis was performed comparing the collected spectra to a high-resolution literature spectrum(Smith et al. 2006) that was convoluted with a 0.75 nm Gaussian function in order to match the resolution of the collection spectrometer. An initial estimate of the number density of the formaldehyde in the absorption cell, $N_{\mathrm{CH_2O}}^{\mathrm{AbsCell}}$ (cm$^{-3}$), was then used to convert the measured absorbance, $A$, to absorption cross-section, $\sigma$. A least-squared minimization analysis was then performed by varying the estimated $N_{\mathrm{CH_2O}}^{\mathrm{AbsCell}}$ to obtain a best fit to the convoluted literature spectrum. Using the $N_{\mathrm{CH_2O}}^{\mathrm{AbsCell}}$ value that gave the best fit, together with the density in the absorption cell ($N_{\mathrm{Total}}^{\mathrm{AbsCell}}$, determined from the total pressure in the cell), the fraction of formaldehyde in the cell could be calculated, and this value then adopted as the fraction of formaldehyde in the low temperature flows generated by the Laval nozzles. The statistical error in the fitted $N_{\mathrm{CH_2O}}^{\mathrm{AbsCell}}$ values was typically around 2 %, significantly smaller than the ~ 10 % uncertainties in the density of the Laval flows.

**CH2O dimerization experiments**

Experiments were conducted monitoring the LIF from $CH_2O$ as a function of $[CH_2O]$ added to the flows. In regions in which little or no $CH_2O$ dimerization occurred, the amount of $CH_2O$ monomer present in the flows would increase linearly with the $[CH_2O]$ added, and as such the $CH_2O$ LIF signal would increase linearly; however, at $CH_2O$ concentrations at which significant dimerization was occurring, the amount of $CH_2O$ monomer present in the flows would actually be less than the $[CH_2O]$ added, and as such the $CH_2O$ LIF signal would be less than expected. Thus by plotting the $CH_2O$ LIF signal *vs* $[CH_2O]$ added to the flows, the point at which the plot begins to curve over indicates the point at which dimers are beginning to form. We found that significant complex formation occurred at much lower $[CH_2O]$ when using either Ar or $N_2$ as a bath gas as compared to He, presumably due to both Ar and $N_2$ being better third bodies than He. In the He flow ~ 34 K, formaldehyde dimerization occurred around a concentration of ~ $3 \times 10^{14}$ molecule cm$^{-3}$, whereas in the $N_2$ flow at ~ 72 K, and the Ar flow at ~ 40 K, dimerization occurred at around $5 \times 10^{13}$ and $3 \times 10^{13}$ molecule cm$^{-3}$ respectively. Dimerization experiments in He were carried out both with and without $CH_4$ and $NH_3$ present, and no change in the concentration at which significant $CH_2O$ dimerization occurred was observed.

**Further details on MESMER calculations**

The full $NH_2$ + $CH_2O$ PE surface (Figure 2) was employed in the MESMER calculations. The inverse Laplace transform parameters for the initial association reaction of $NH_2$ with $CH_2O$, which take the form of a modified Arrhenius function ($k = (A/298)^n$), were $A = 1.89 \times 10^{-11}$ cm$^3$ molecule$^{-1}$ s$^{-1}$ and $n$ = -0.94. This means that at $T$ < 50 K, the association rate is > $1 \times 10^{-10}$ cm$^3$ molecule$^{-1}$ s$^{-1}$, consistent with low temperature capture rates. These $A$ and $n$ values were obtained from fitting a calculated $NH_2$ + NO PES to experimental rate coefficients over the temperature range 35 – 2500 K, the results from which will be published in a separate paper. The exponential down model was used to estimate the probability of collisional transfer between grains. For $N_2$ as the third body, the average energy for downward transitions ($\Delta E_{down}$) was set to 250 cm$^{-1}$ at 298 K, with a temperature dependence of $T^{0.25}$. Rate coefficients and branching ratios (BRs) were determined over a range of third body pressures. The low-pressure rate coefficients and BRs (Table S3) were determined by lowering the third body pressure until the calculated rate coefficients and BRs were effectively constant; this was achieved at a pressure of $[N_2]$ = $1 \times 10^{14}$ molecule cm$^{-3}$ (see Figure S4).

**Table S1.** Molecular properties of the stationary points on the potential energy surface for $NH_2$ + $CH_2O$ calculated at the M062X/aug-cc-pVTZ level of theory.

| Molecule | Geometries (Cartesian coordinates in Å) | Rotational Constants (cm$^{-1}$) | Unscaled Vibrational Frequencies (cm$^{-1}$) |
|---|---|---|---|
| $NH_2$ | N, -1.421509, 0.178069, 0.816497<br>H, -1.05166, 0.701101, 1.623253<br>H, -1.05166, 0.701101, 0.009741 | 23.322, 12.854, 8.287 | 1515, 3390, 3482 |
| $CH_2O$ | C, -0.839414, -0.559066, 1.840409<br>H, -0.799038, -1.657637, 1.932365<br>H, -0.477756, -0.130005, 0.890643<br>O, -1.253888, 0.131206, 2.724836 | 9.501, 1.316, 1.156 | 1215, 1274, 1539, 1868, 2947, 3016 |
| HB-PRC | C, -3.935826, -0.136227, -0.685981<br>H, -4.264535, -0.504606, -1.669994<br>H, -3.949993, -0.847361, 0.156089<br>O, -3.574759, 0.998092, -0.533267<br>N, -4.159377, 1.256065, -3.511158<br>H, -3.843663, 1.691567, -2.637281<br>H, -4.088938, 2.001697, -4.208166 | 1.261, 0.168, 0.148 | 109.2543, 149.1058, 162.6028, 172.2076, 227.8815, 272.2967, 1231.6183, 1286.8207, 1533.3299, 1541.0208, 1851.4443, 2964.3051, 3052.2601, 3396.2545, 3490.5585 |
| TS1 (from HB-PRC to $NH_3$ + CHO) | C, -3.831573, 0.352703, -0.872888<br>H, -4.143528, 0.889932, -1.931119<br>H, -4.657147, 0.281562, -0.135048<br>O, -2.736788, -0.04624, -0.662768<br>N, -4.393867, 1.385868, -3.243245<br>H, -3.421053, 1.360913, -3.563734<br>H, -4.801601, 0.538062, -3.648931 | 1.655, 0.168, 0.156 | 1469.8414$i$, 61.5819, 132.3699, 305.6244, 642.2758, 705.5153, 824.808, 1215.9859, 1287.3805, 1491.1897, 1536.0884, 1904.3624, 2905.943, 3407.1285, 3496.9579 |
| $NH_3$ | N, -0.771215, 0.655501, 2.262482<br>H, -0.396641, -0.284739, 2.26248<br>H, -0.39663, 1.125585, 3.076759<br>H, -0.396639, 1.125586, 1.448201 | 10.006, 10.005, 6.309 | 1031.9367, 1659.05, 1659.6368, 152.1677, 16.8113, 16.8893 |
| CHO | C, -0.873151, 0.590075, 2.371278<br>O, 0.191464, 0.476858, 2.837829<br>H, -1.282856, -0.072704, 1.566428 | 24.057, 1.516, 1.426 | 1100.2964, 1993.569, 2728.3276 |
| vW-PRC | C, -3.794329, 0.152395, -0.811276<br>H, -3.795422, -0.84753, -1.273505<br>H, -4.774027, 0.541528, -0.491533<br>O, -2.784243, 0.780725, -0.663402<br>N, -4.469033, 1.106528, -3.351236<br>H, -3.605626, 1.631134, -3.179343<br>H, -4.762878, 1.398019, -4.287436 | 1.201, 0.174, 0.157 | 56.5058, 87.5845, 156.5873, 203.4769, 243.5292, 266.7348, 1211.3078, 1273.8319, 1528.1523, 1542.5043, 1857.4736, 2965.0126, 3036.7861, 3405.915, 3498.8421 |
| TS2 (from vW-PRC to Adduct) | C, -2.26331, 0.183381, 0.117246<br>O, -2.482898, -0.503947, -0.876099<br>H, -2.910951, 0.130668, 1.002726<br>H, -1.289447, 0.66661, 0.273345<br>N, -3.139859, 1.84904, -0.606732<br>H, -2.654415, 1.959102, -1.501061<br>H, -4.063863, 1.49354, -0.866317 | 1.094, 0.311, 0.265 | 376.2335$i$, 227.3008, 338.3713, 605.1296, 762.5614, 775.3409, 1171.4991, 1248.9821, 1497.5487, 1544.5383, 1672.2704, 2986.8724, 3056.6958, 3418.2836, 3510.6696 |
| Adduct | C, -3.53812, -0.413349, 0.081098<br>H, -3.042506, -0.166352, 1.031799<br>H, -3.416556, -1.500674, -0.034757<br>N, -2.874863, 0.263774, -0.998607<br>H, -3.290149, 0.019246, -1.888234<br>H, -2.940992, 1.267606, -0.890828<br>O, -4.873764, -0.186883, 0.266368 | 1.496, 0.339, 0.305 | 271.9861, 506.4662, 699.9390, 837.3067, 1018.8172, 1027.9589, 1144.2063, 1347.2773, 1348.2099, 1403.6108, 1659.9952, 2983.2752, 3007.3602, 3541.0056, 3624.8953 |
| TS3 (from Adduct to Formamide + H) | C, -3.828356, -0.40751, -0.097158<br>H, -2.710929, -0.393618, 1.04196<br>H, -3.648169, -1.492848, -0.174997<br>N, -3.275334, 0.295334, -1.14695<br>H, -2.439768, -0.066727, -1.573964<br>H, -3.350992, 1.299259, -1.095917<br>O, -4.7234, 0.049476, 0.613865 | 1.629, 0.357, 0.319 | 1103.7717$i$, 408.7874, 485.0182, 530.0242, 604.7571, 648.3972, 1058.1238, 1129.4240, 1250.5444, 1380.2986, 1608.7231, 1673.9403, 2974.6377, 3588.2631, 3715.8247 |
| Formamide | C, -0.294319, -0.062605, 2.242845<br>H, -0.516772, -0.437552, 1.231363<br>O, -0.839206, 0.905013, 2.712192<br>N, 0.643736, -0.802983, 2.881341<br>H, 1.067144, -1.599702, 2.443436<br>H, 0.921033, -0.544525, 3.81346 | 2.461, 0.383, 0.331 | 226.5062, 572.2045, 635.9463, 1060.5635, 1063.3112, 1279.6132, 1427.7140, 1613.7487, 1834.7948, 3010.2805, 3609.8168, 3753.1700 |

**Table S2.** Heats of reaction and barrier heights for the $NH_2 + CH_2O$ reaction.

| Molecule | Relative Energies (ZPE corrected; kJ mol$^{-1}$) | | | | | Relative Electronic Energies (no ZPE correction; kJ mol$^{-1}$) | | | |
|---|---|---|---|---|---|---|---|---|---|
| | This Study[a] | Li and Lü[b] | Barone[c] | Vazart[d] | Song[e] | This Study[a] | Barone[c] | Vazart[d] | Song[e] |
| HB-PRC | -9.06 | | | | | -15.87 | -15.9[e] | | |
| TS1 (from HB-PRC to $NH_3$ + CHO) | 23.14 | 24.64 | | | | 25.05 | | | |
| $NH_3$ + CHO (R1a) | -79.33 | -77.57 | | | | -83.32 | | | |
| vW-PRC | -5.82 | | | | | -12.01 | | -12.2 | |
| TS2 (from vW-PRC to Adduct) | 18.35 | | | 18.9 | 17.8 | 3.68 | | 3.6 | 2.7 |
| Adduct | -48.32 | | -46.9 | -49.8 | | -72.17 | -71.5 | -74.5 | |
| TS3 (from Adduct to Formamide + H) | 5.24 | | 0.2 | 2.9 | | 0.63 | -5 | -1.3 | |
| Formamide + H (R1b) | -44.92 | | -48.5 | -46.9 | | -43.99 | -47.3 | -46.0 | |

[a] Structures and ZPEs calculated at the M062X/aug-cc-PVTZ level of theory. Single point energies calculated at the CCSD(T)/aug-cc-pVTZ+QZ+5Z level of theory, and extrapolated to the complete basis set limit using a mixed Gaussian/exponential extrapolation scheme as proposed by Peterson *et. al.*(Peterson et al. 1994)

[b] Calculated at the G2// UMP/G-311+G(d,p) level.(Li and Lü 2002)

[c] CBS-QB3 electronic energies with B2PLYP-D3/m-aug-cc-pVTZ ZPEs.(Barone et al. 2015)

[d] CCSD(T)/CBS+CV electronic energies with B2PLYP-D3/m-aug-cc-pVTZ ZPEs.(Vazart et al. 2016)

[e] Structures optimized at the M062X/def2-TZVP level and single point energies calculated at the UCCSD(T)-F12/cc-pVTZ-F12 level using NWCHEM.(Song and Kästner 2016)

**Table S3.** Low-pressure limiting rate coefficients and BRs for the reaction between $NH_2$ and $CH_2O$ as predicted by MESMER. Values determined by reducing the pressure of the third body ($N_2$) in the MESMER calculations until the rate coefficients and BRs were effectively constant. See Figure S4 for pressure dependence.

| T / K | $k(NH_2 + CH_2O)$ / $cm^3$ molecule$^{-1}$ s$^{-1}$ | % BR for formamide + H production | T / K | $k(NH_2 + CH_2O)$ / $cm^3$ molecule$^{-1}$ s$^{-1}$ | % BR for formamide + H production |
|---|---|---|---|---|---|
| 10 | 5.19E-12 | 0.00% | 180 | 8.14E-16 | 0.27% |
| 15 | 1.26E-12 | 0.00% | 185 | 8.05E-16 | 0.36% |
| 20 | 4.48E-13 | 0.00% | 190 | 8.01E-16 | 0.46% |
| 25 | 2.01E-13 | 0.00% | 195 | 8.01E-16 | 0.59% |
| 30 | 1.05E-13 | 0.00% | 200 | 8.05E-16 | 0.75% |
| 35 | 6.17E-14 | 0.00% | 205 | 8.14E-16 | 0.92% |
| 35 | 7.99E-14 | 0.00% | 210 | 8.28E-16 | 1.12% |
| 40 | 4.85E-14 | 0.00% | 215 | 8.46E-16 | 1.35% |
| 45 | 3.16E-14 | 0.00% | 220 | 8.68E-16 | 1.59% |
| 50 | 2.18E-14 | 0.00% | 225 | 8.96E-16 | 1.86% |
| 55 | 1.56E-14 | 0.00% | 230 | 9.28E-16 | 2.15% |
| 60 | 1.17E-14 | 0.00% | 235 | 9.65E-16 | 2.45% |
| 65 | 8.95E-15 | 0.00% | 240 | 1.01E-15 | 2.77% |
| 70 | 7.05E-15 | 0.00% | 245 | 1.06E-15 | 3.10% |
| 75 | 5.67E-15 | 0.00% | 250 | 1.11E-15 | 3.43% |
| 80 | 4.65E-15 | 0.00% | 255 | 1.17E-15 | 3.77% |
| 85 | 3.88E-15 | 0.00% | 260 | 1.24E-15 | 4.11% |
| 90 | 3.28E-15 | 0.00% | 265 | 1.31E-15 | 4.44% |
| 95 | 2.81E-15 | 0.00% | 270 | 1.39E-15 | 4.77% |
| 100 | 2.44E-15 | 0.00% | 275 | 1.48E-15 | 5.09% |
| 105 | 2.15E-15 | 0.00% | 280 | 1.58E-15 | 5.41% |
| 110 | 1.90E-15 | 0.00% | 285 | 1.68E-15 | 5.71% |
| 115 | 1.71E-15 | 0.00% | 290 | 1.80E-15 | 6.00% |
| 120 | 1.54E-15 | 0.00% | 295 | 1.92E-15 | 6.27% |
| 125 | 1.41E-15 | 0.00% | 300 | 2.06E-15 | 6.53% |
| 130 | 1.29E-15 | 0.00% | 305 | 2.20E-15 | 6.77% |
| 135 | 1.20E-15 | 0.01% | 310 | 2.36E-15 | 7.00% |
| 140 | 1.12E-15 | 0.01% | 315 | 2.53E-15 | 7.21% |
| 145 | 1.05E-15 | 0.02% | 320 | 2.71E-15 | 7.41% |
| 150 | 9.92E-16 | 0.03% | 325 | 2.91E-15 | 7.59% |
| 155 | 9.45E-16 | 0.05% | 330 | 3.12E-15 | 7.76% |
| 160 | 9.06E-16 | 0.07% | 335 | 3.34E-15 | 7.91% |
| 165 | 8.74E-16 | 0.10% | 340 | 3.58E-15 | 8.05% |
| 170 | 8.48E-16 | 0.14% | 345 | 3.83E-15 | 8.17% |
| 175 | 8.29E-16 | 0.20% | 350 | 4.10E-15 | 8.28% |

**Figure S4.** Temperature and pressure dependent rate coefficients and BRs for the reaction between $NH_2 + CH_2O$ as predicted by MESMER. Top panel: the overall rate coefficient, $k_1$. Bottom panel: the branching ratio to formamide ($NH_2CHO$). Both are plotted as a function of $[N_2]$ and $T$.

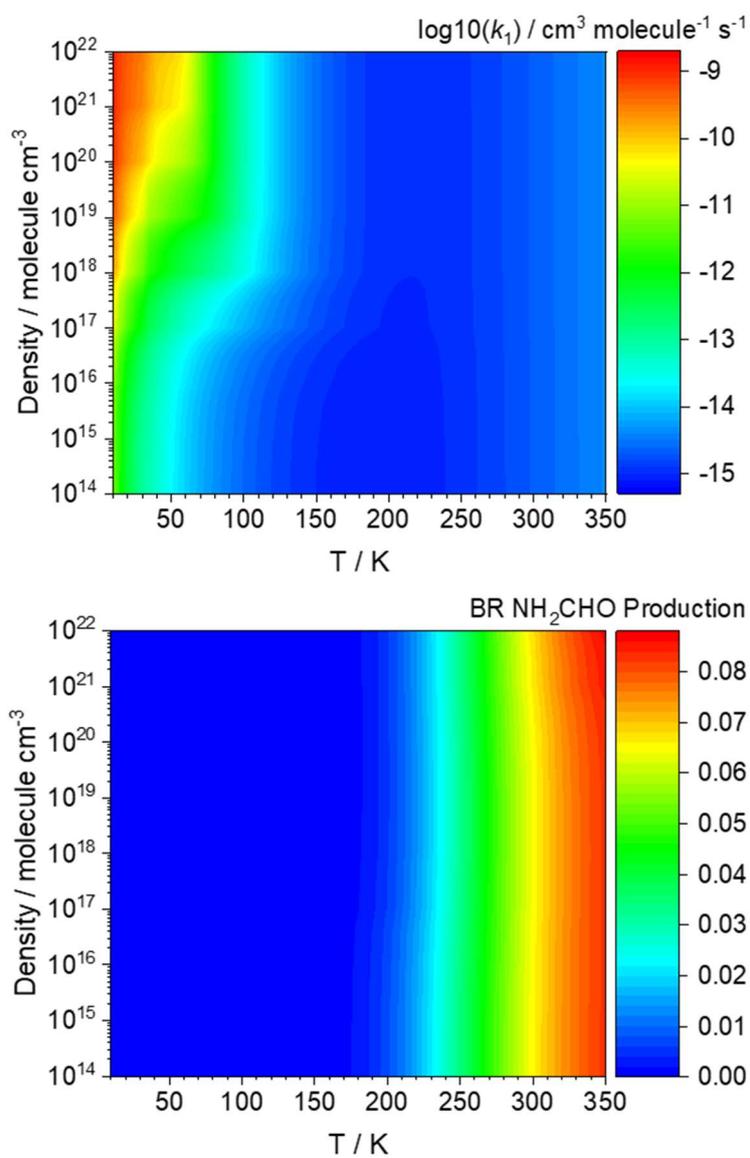

**Figure S5.** Predicted rate coefficients at a total pressure of $[N_2] = 1 \times 10^{17}$ molecule cm$^{-3}$ for the H-abstraction (black squares) and formamide + H (red circles) product channels of the $NH_2$ + $CH_2O$ reaction. The experimentally determined upper limit is shown as a yellow line with a downward arrow below to indicate that the predicted is significantly smaller than this.

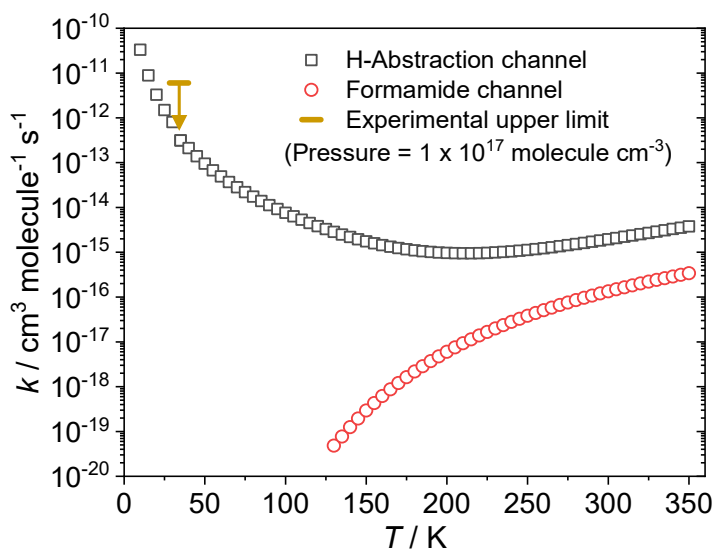

**References.**